# Generalization of Bidirectional Controlled Hybrid Protocol of Quantum Teleportation and Remote State Preparation

MohammadHossein Valeh[1], Hossein Aghababa[2,3,a], Mohammadreza Kolahdouz[1], Masoud Barati[4]

[1]School of Electrical and Computer Engineering, Faculty of Engineering, University of Tehran, Tehran, Iran

[2]Faculty of Engineering, College of Farabi, University of Tehran, Tehran, Iran

[3]Department of Engineering, Loyola University Maryland, Maryland, USA

[4]Swanson School of Engineering, Electrical Engineering, University of Pittsburgh, Pittsburgh, PA, USA

[a]haghababa@loyola.edu

**Abstract**

This paper proposes a protocol for hybrid quantum teleportation and remote state preparation(RSP). It allows users to transmit an n-qubit quantum state by using (4n+1)qubits entangled state as a quantum channel to each other simultaneously. This protocol is under the control of a third person, Charlie, who ensures the security of this transmission.

## 1. Introduction

Quantum computing is gaining prominence today due to its potential to solve complex problems beyond the capabilities of classical computers. Quantum computing is enhancing communication by enabling ultra-secure data transmission. It also optimizes network performance by analyzing complex data for efficient configurations and resource allocation. Additionally, quantum algorithms improve error correction and noise reduction, leading to more reliable communication channels. Quantum teleportation and remote state preparation (RSP) are essential to modern communication technologies, enabling the transfer of quantum information with enhanced security and efficiency[1]. Quantum teleportation allows for transmitting unknown quantum states between distant parties without physical transfer, ensuring secure information exchange. RSP facilitates the preparation of a known quantum state at a remote location using pre-shared entanglement and minimal classical communication, optimizing resource use. These techniques are foundational for developing advanced quantum networks and secure communication protocols[2].

In 1993, Charles H. Bennett and his colleagues introduced the concept of quantum teleportation; this process utilizes a combination of Einstein-Podolsky-Rosen (EPR) correlations [3], and some complementary works have been done in recent years, such as introducing bidirectional teleportation [4-7]. These works included the introduction of a bidirectional protocol in both single-qubit and multi-qubit forms. After that, controlled quantum teleportation was introduced, where a third party was involved in the communication, whose role was to ensure the security of the communication. These works also were in the form of single-qubit and multi-qubits[8-12]. The concept of RSP was introduced as a variant of quantum teleportation, where the sender knows the quantum state to be communicated. In traditional quantum teleportation, an unknown quantum state is transmitted using shared entanglement and classical communication. In contrast, RSP utilizes the sender's knowledge of the state to reduce the classical communication required [13]. Over the years, RSP has been explored in various contexts, including its application in quantum networks and its implementation in different physical systems. For instance, recent research has investigated the use of RSP in genuine quantum networks, highlighting its potential for efficient state preparation in distributed quantum systems [14]. One study was performed in 2015, and they proposed the concept of bidirectional controlled joint remote state preparation (BCJRSP), which essentially fuses the ideas of bidirectional controlled teleportation and joint remote state preparation [15]. Hybrid quantum communication protocols that combine quantum teleportation and remote state preparation (RSP) have been proposed to enhance the efficiency and flexibility of quantum information transfer. For instance, a six-party scheme implements quantum teleportation and joint RSP under a controller's supervision, utilizing a multi-qubit entangled state as the quantum channel, achieving a 100% success probability [16]. Another protocol employs a seven-qubit entangled state to facilitate controlled bidirectional communication, allowing one party to teleport an arbitrary single-qubit state to another while simultaneously enabling the latter to remotely prepare a known state for the former, all under the supervision of controllers [17]. Some of these studies have introduced and designed quantum teleportation protocols for n-qubit systems [18]. For example, in 2022, Mafi, Y. et al. introduced bidirectional quantum teleportation for a system that utilized a 2n-qubit quantum channel to transfer quantum states with an arbitrary number of qubits [19]. In another study, A new protocol of bidirectional quantum teleportation (BQT) is proposed in which the users can transmit a class of n-qubit states to each other simultaneously by using (2n + 2)-qubit entangled states as a quantum channel[20]. In some other research, the use of cluster states for implementing quantum channels has been investigated [21-24]. Also, some studies have investigated quantum teleportation through a noisy quantum channel [25-27]. In the most recent study, controlled bidirectional quantum teleportation was combined with remote state preparation in a hybrid manner. This system was designed for two qubits Alice teleport her state to Bob, and Bob simultaneously prepares his state remotely in Alice's location [28-30].

This paper aims to introduce a controlled bidirectional protocol in which an n-qubit state located with Alice is to be teleported to Bob through a quantum channel. Simultaneously, Bob will remotely prepare a known n-qubit quantum state, which he possesses, at Alice's location. This communication is controlled by Charlie, whose role is to ensure the security of the process.

## 2. The proposed protocol

In this section, we'll discuss our proposed scheme where Alice teleports her state to Bob, and Bob simultaneously prepares his state remotely at Alice's location. Let's assume that Alice has an unknown n-qubit state as follows:

$$|\chi\rangle_{a_1 a_2 ... a_n} = \sum_{i=0}^{2^n - 1} \alpha_i |i\rangle \quad (1)$$

And Bob also has a known n-qubit state as follows:

$$|\varphi\rangle_{b_1 b_2 ... b_n} = \sum_{j=0}^{2^n - 1} \beta_j e^{i\theta_j} |j\rangle \quad (2)$$

Where the following relationship exists between coefficients:

$$\sum_{i=0}^{2^n - 1} |\alpha_i|^2 = 1, \sum_{j=0}^{2^n - 1} |\beta_j|^2 = 1 \quad (3)$$

There is also an entangled (4n+1)qubit channel shared between Alice, Bob, and Charlie. Of these qubits, 2n qubits are assigned to Alice, which is form $A_1$ to $A_{2n}$; 2n qubits are assigned to Bob, which is form $B_1$ to $B_{2n}$; and one qubit C is assigned to Charlie, the third party who is supposed to ensure the security of this communication. This quantum channel consists of the entanglement of Alice, Bob, and Charlie qubits, which are placed in the initial state of the Bell bases. To create this entanglement, we use a tensor product between these quantum states. The quantum channel is in the form of this:

$$|\psi\rangle_{CH} = \frac{1}{\sqrt{2}} \left( |\varphi^+\rangle_{A_1 B_1} |\varphi^+\rangle_{A_2 B_2} ... |\varphi^+\rangle_{A_{2n} B_{2n}} |0\rangle_C + |\psi^-\rangle_{A_1 B_1} |\psi^-\rangle_{A_2 B_2} ... |\psi^-\rangle_{A_{2n} B_{2n}} |1\rangle_C \right) \quad (4)$$

where $|\varphi^+\rangle = \frac{1}{\sqrt{2}}(|00\rangle + |11\rangle)$, $|\psi^-\rangle = \frac{1}{\sqrt{2}}(|01\rangle - |10\rangle)$.

A tensor product must be applied to the system's inputs to obtain its quantum state. Therefore, the overall quantum state of the system is obtained by using the tensor product between the quantum states of Alice, Bob, and the channel, as follows.

$$|\psi\rangle_s = |\chi\rangle_{a_1 ... a_n} \otimes |\psi\rangle_{CH}$$
$$= (\alpha_0 |0\rangle + ... + \alpha_{2^n - 1} |2^n - 1\rangle)_{a_1 ... a_n} \quad (5)$$
$$\times \frac{1}{\sqrt{2}} \left( |\varphi^+\rangle_{A_1 B_1} |\varphi^+\rangle_{A_2 B_2} ... |\varphi^+\rangle_{A_{2n} B_{2n}} |0\rangle_C + |\psi^-\rangle_{A_1 B_1} |\psi^-\rangle_{A_2 B_2} ... |\psi^-\rangle_{A_{2n} B_{2n}} |1\rangle_C \right)$$

This protocol is carried out in six steps, which we will examine in the following.

1- At first, Alice performs a Bell-state measurement on her particles, which include $(A_1, a_1) ... (A_n, a_n)$, and informs Bob about the result.

The Bell bases for this system are defined as follows. In this context, the value of k can be any number smaller than n, and $a_k A_k$ refers to the pair of qubits on which the measurement is performed together. The schematic of this protocol and the process of its operations are illustrated in Figure 1.

$$|\varphi^+\rangle_{a_k A_k} = \frac{1}{\sqrt{2}}(|00\rangle + |11\rangle)_{a_k A_k},$$

$$|\varphi^-\rangle_{a_k A_k} = \frac{1}{\sqrt{2}}(|00\rangle - |11\rangle)_{a_k A_k},$$

$$|\psi^+\rangle_{a_k A_k} = \frac{1}{\sqrt{2}}(|01\rangle + |10\rangle)_{a_k A_k}, \quad (6)$$

$$|\psi^-\rangle_{a_k A_k} = \frac{1}{\sqrt{2}}(|01\rangle - |10\rangle)_{a_k A_k}$$

We know that in systems where qubits are entangled, measuring one of the qubits causes the remaining qubits to collapse. Given that Alice's measurement of her qubits is performed using the Bell basis, $4^n$ possible outcomes may be obtained. For simplicity in proving this scheme, we assume, for example, that Alice's measurement result is equal to $|\varphi^+\rangle_{a_1 A_1}...|\varphi^+\rangle_{a_n A_n}$. The collapsed state of the remaining qubits is in the form.

$$|\psi^1\rangle = \langle \varphi^+_{a_m A_n}|\langle \varphi^+_{a_m A_m}|\langle \varphi^+|\psi\rangle_s$$

$$= \frac{1}{\sqrt{2}}|\varphi^+\rangle_{a_m A_m}|\varphi^+\rangle_{a_n A_n}(|\varphi^+\rangle_{A_1 B_1}|\varphi^+\rangle_{A_2 B_2}...|\varphi^+\rangle_{A_{2n} B_{2n}}|0\rangle_C + |\psi^-\rangle_{A_1 B_1}|\psi^-\rangle_{A_2 B_2}...|\psi^-\rangle_{A_{2n} B_{2n}}|1\rangle_C)$$

$$(\alpha_0|0\rangle + ... + \alpha_{2^n-1}|2^n-1\rangle)_{a_1...a_n} \quad (7)$$

$$= \frac{1}{2\sqrt{2}}\begin{bmatrix}(\alpha_0|0\rangle + ... + \alpha_{2^n-1}|2^n-1\rangle)_{B_1...B_n}(|\varphi^+\rangle...|\varphi^+\rangle)_{A_{n+1}B_{n+1}..A_{2n}B_{2n}}|0\rangle_C \\ +(\alpha_{2^n-1}|0\rangle + ... + \alpha_0|2^n-1\rangle)_{B_1...B_n}(|\psi^-\rangle...|\psi^-\rangle)_{A_{n+1}B_{n+1}..A_{2n}B_{2n}}|1\rangle_C\end{bmatrix}$$

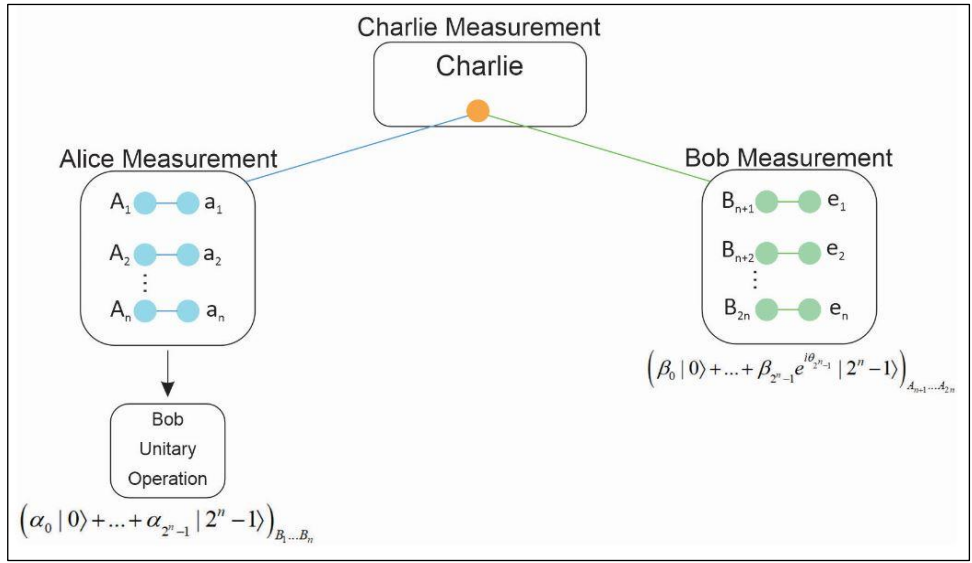

Figure 1-Schematic of proposed model

2- To perform the C-not operation that Bob will carry out in the next step, Bob must introduce several auxiliary qubits to facilitate the operation. The introduction of these qubits mainly serves as a control mechanism. At this stage, Bob introduces n auxiliary qubits, denoted as $e$, which are initially in the $|0\rangle$ state. These qubits are entangled with the qubits ($B_{n+1}\ldots B_{2n}$). With the introduction of these auxiliary qubits, the overall quantum state of the system changes as follows.

$$|\psi^1\rangle = \frac{1}{2\sqrt{2}}\begin{bmatrix}(\alpha_0|0\rangle+\ldots+\alpha_{2^n-1}|2^n-1\rangle)_{B_1\ldots B_n}(|\varphi^+\rangle\ldots|\varphi^+\rangle)_{A_{n+1}B_{n+1}\ldots A_{2n}B_{2n}}|0\ldots 0\rangle_{e_1\ldots e_n}|0\rangle_C \\ +(\alpha_{2^n-1}|0\rangle+\ldots+\alpha_0|2^n-1\rangle)_{B_1\ldots B_n}(|\psi^-\rangle\ldots|\psi^-\rangle)_{A_{n+1}B_{n+1}\ldots A_{2n}B_{2n}}|0\ldots 0\rangle_{e_1\ldots e_n}|1\rangle_C\end{bmatrix} \quad (8)$$

3- In this step, Bob performs a C-not operation on qubits ($B_{n+1}, e_1$)…($B_{2n}, e_n$). This C-not operation is performed with control qubit B and target qubit $e$. The state after this operation is in the form below.

$$|\psi'^1\rangle = \frac{1}{2\sqrt{2}}\begin{bmatrix}(\alpha_0|0\rangle+\ldots+\alpha_{2^n-1}|2^n-1\rangle)_{B_1\ldots B_n}\frac{1}{\sqrt{2}}(|000\rangle+\ldots+|111\rangle)_{A_{n+1}B_{n+1}e_1}\ldots(|000\rangle+\ldots+|111\rangle)_{A_{2n}B_{2n}e_n}|0\rangle_C \\ +(\alpha_{2^n-1}|0\rangle+\ldots+\alpha_0|2^n-1\rangle)_{B_1\ldots B_n}\frac{1}{\sqrt{2}}(|011\rangle+\ldots+|100\rangle)_{A_{n+1}B_{n+1}e_1}\ldots(|011\rangle+\ldots+|100\rangle)_{A_{2n}B_{2n}e_n}|1\rangle_C\end{bmatrix}$$

(9)

4- In this step Bob performs a measurement on qubits ($B_{n+1}\ldots B_{2n}$). This measurement is performed based on the following orthogonal bases, which correspond to the coefficients of Bob's initial state.

$$|\varphi^1\rangle_B = (\beta_0|0\rangle + \beta_1|1\rangle)_B,$$
$$|\varphi^2\rangle_B = (\beta_1|0\rangle - \beta_0|1\rangle)_B \quad (10)$$

After this measurement is performed, the remaining qubits collapse. Assuming that the measurement result at this step is $|\varphi^1\rangle$, the collapsed state can be written as follows.

$$|\psi'^{1,1}\rangle = {}_{B_{n+1}\ldots B_{2n}}\langle\varphi^1|\psi'^1\rangle$$
$$= \frac{1}{4\sqrt{2}}\begin{bmatrix}(\alpha_0|0\rangle+\ldots+\alpha_{2^n-1}|2^n-1\rangle)_{B_1\ldots B_n}(\beta_0|0\rangle|0\rangle+\ldots+\beta_{2^n-1}|2^n-1\rangle|0\rangle)_{A_{n+1}e_1\ldots A_{2n}e_n}|0\rangle_C \\ +(\alpha_{2^n-1}|0\rangle+\ldots+\alpha_0|2^n-1\rangle)_{B_1\ldots B_n}(\beta_{2^n-1}|0\rangle+\ldots+\beta_0|2^n-1\rangle)_{A_{n+1}e_1\ldots A_{2n}e_n}|1\rangle_C\end{bmatrix} \quad (11)$$

5- At this stage, Bob needs to perform a measurement on the auxiliary qubits ($e_1\ldots e_n$). The basis for this measurement depends on the result obtained from the measurement in the previous step. For example, if the measurement result on qubit B is $|\varphi^1\rangle_B$, then the following orthogonal bases are considered for measuring the qubit $e$.

$$|\varphi^1\rangle_e = \frac{1}{\sqrt{2}}\left(|0\rangle + e^{-i\theta}|1\rangle\right)_e,$$
$$|\varphi^2\rangle_e = \frac{1}{\sqrt{2}}\left(|0\rangle - e^{-i\theta}|1\rangle\right)_e \tag{12}$$

But if the measurement result of qubit B is equal to $|\varphi^2\rangle_B$, the measurement of qubit $e$ is done with the following bases.

$$|\varphi^{1'}\rangle_e = \frac{1}{\sqrt{2}}\left(e^{-i\theta}|0\rangle + |1\rangle\right)_e,$$
$$|\varphi^{2'}\rangle_e = \frac{1}{\sqrt{2}}\left(e^{-i\theta}|0\rangle - |1\rangle\right)_e \tag{13}$$

Here, we assume that the result of this measurement is equal to $|\varphi^1\rangle$, in this case, after measuring the auxiliary qubit, the quantum state of the remaining collapsed qubits is written as follows.

$$|\psi'^{1,1,1}\rangle = {}_{e_1...e_n}\langle\varphi^1|\psi'^{1,1}\rangle$$
$$= \frac{1}{4\sqrt{2}}\left[\begin{array}{l}(\alpha_0|0\rangle + ... + \alpha_{2^n-1}|2^n-1\rangle)_{B_1...B_n}\left(\beta_0|0\rangle + ... + \beta_{2^n-1}e^{i\theta_{2^n-1}}|2^n-1\rangle\right)_{A_{n+1}...A_{2n}}|0\rangle_C \\ +(\alpha_{2^n-1}|0\rangle + ... + \alpha_0|2^n-1\rangle)_{B_1...B_n}\left(\beta_{2^n-1}e^{i\theta_{2^n-1}}|0\rangle + ... + \beta_0|2^n-1\rangle\right)_{A_{n+1}...A_{2n}}|1\rangle_C\end{array}\right] \tag{14}$$

6- In the next step, Charlie performs a von Neumann measurement on qubit C, which is under his assignment. As mentioned earlier, Charlie's role is primarily as the controller of this communication. Therefore, it is only necessary for Charlie to perform a single measurement on the qubits to verify the integrity of the communication. Since this qubit is a single bit, the measurement is conducted in the basis $|0\rangle$ and $|1\rangle$. After performing this measurement, we assume, for example, that the measurement result is $|0\rangle$. With this assumption, the collapsed quantum state of the remaining qubits is obtained as follows.

$$|\psi'^{1,1,1,0}\rangle = {}_C\langle 0|\psi'^{1,1,1}\rangle$$
$$= \frac{1}{4\sqrt{2}}(\alpha_0|0\rangle + ... + \alpha_{2^n-1}|2^n-1\rangle)_{B_1...B_n}\left(\beta_0|0\rangle + ... + \beta_{2^n-1}e^{i\theta_{2^n-1}}|2^n-1\rangle\right)_{A_{n+1}...A_{2n}} \tag{15}$$

At this stage, we observe that the coefficients of the initial quantum states of Alice and Bob have been transferred to one another. It can be concluded that the controlled bidirectional quantum teleportation, designed as a hybrid with remote state preparation, is successfully performed. The steps of this protocol is described in Figure 2.

As mentioned at the beginning of the protocol's introduction, for simplicity, we considered a specific case of Alice's measurement result and used it as the basis for introducing and describing the protocol's steps. The coefficients of Alice's quantum state, which are transferred to Bob's qubits in Step 1, vary depending on the outcome of Alice's measurement. In this case, an appropriate unitary operation must be applied to restore Alice's original quantum state. This unitary operation is performed on Bob's qubits.

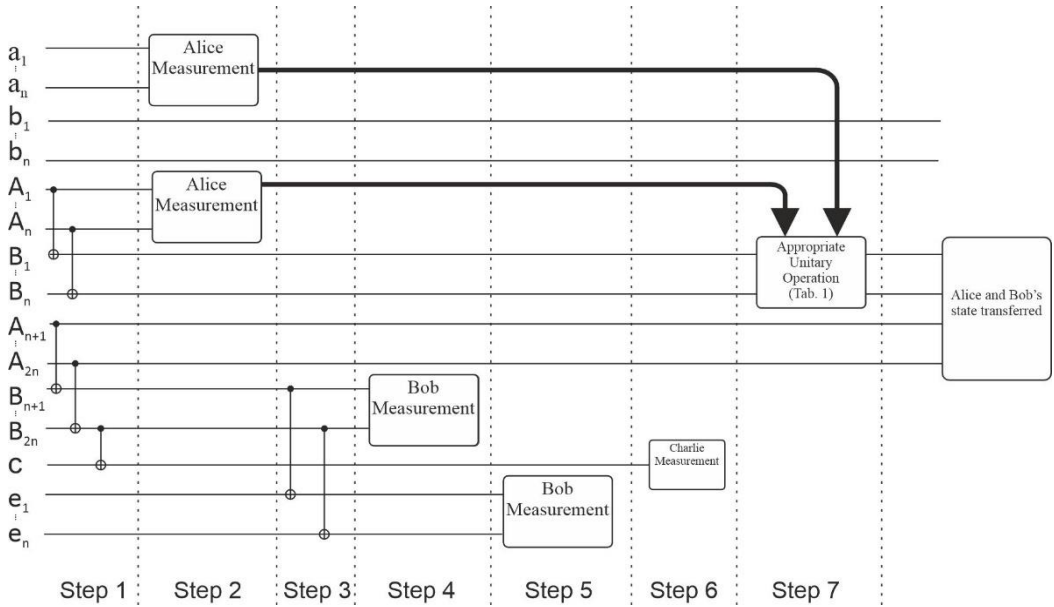

Figure 2-Steps of protocol

The choice of the unitary operator depends on Alice's measurement in Step 1 and Charlie's measurement in the final step. Table 1 presents the suitable unitary operators for restoring Alice's quantum state for each combination of Alice's and Charlie's measurement outcomes.

Table 1-Suitable Unitary Operator

| Alice's measurement | Charlie's measurement | Unitary operation |
| --- | --- | --- |
| $|\varphi^+\rangle_{a_k A_k}$ | $|0\rangle$ | $I$ |
| $|\varphi^-\rangle_{a_k A_k}$ | $|0\rangle$ | $Z$ |
| $|\psi^+\rangle_{a_k A_k}$ | $|0\rangle$ | $X$ |
| $|\psi^-\rangle_{a_k A_k}$ | $|0\rangle$ | $-XZ$ |
| $|\varphi^+\rangle_{a_k A_k}$ | $|1\rangle$ | $-XZ$ |
| $|\varphi^-\rangle_{a_k A_k}$ | $|1\rangle$ | $X$ |
| $|\psi^+\rangle_{a_k A_k}$ | $|1\rangle$ | $-Z$ |
| $|\psi^-\rangle_{a_k A_k}$ | $|1\rangle$ | $-I$ |

According to this table, for each measurement result of the qubit pairs measured by Alice and the measurement result of Charlie's qubit, the corresponding unitary operator is selected and applied to Bob's qubits. For scenarios where the system involves two or more qubits, additional qubits are measured in the first step, resulting in measurement outcomes comprising multiple Bell bases. In such cases, for each measurement outcome, the appropriate unitary operator is selected, and a tensor product is performed among them. Finally, the resulting unitary operator is applied to all of Bob's qubits.

For example, suppose that in a two-qubit system, the result of step 1 measurement for qubits a₁A₁ and a₂A₂ is obtained as follows.

$$|\psi^-\rangle_{a_1 A_1} |\varphi^+\rangle_{a_2 A_2} \tag{16}$$

If Charlie's measurement result in the last step is equal to $|0\rangle$, the unitary operator required to reconstruct Alice's state is defined as follows.

$$U = (-XZ) \otimes (I) \tag{17}$$

Also, if Charlie's measurement result is equal to $|1\rangle$, we will introduce the unitary operator as follows.

$$U = (-I) \otimes (-XZ) \tag{18}$$

In this part, we examined the mathematical model of the proposed protocol. We saw that by performing appropriate operations and measurements, we can transfer Alice and Bob's quantum states to each other.

## 3. Comparison

One of the proposed protocol's main advantages is that users can simultaneously transmit n qubits to one another. This contrasts with previously introduced protocols, where only up to two quantum states could be transmitted through teleportation or remote state preparation. Additionally, as mentioned earlier, this protocol requires a two-qubit Bell-state measurement. This is advantageous because it simplifies experimental implementation and makes it more practical.

Another significant and key advantage of this protocol is that the communication between Alice and Bob relies solely on a quantum channel with a specific number of qubits. In contrast, other protocols typically require both quantum and classical channels to transmit information. The proposed model offers certain advantages while also presenting some drawbacks, such as greater implementation complexity. However, with advancements in quantum computing hardware, it can be expected that some of this implementation complexity will be reduced.

To measure the efficiency of this protocol, we use the following relation.

$$\eta = \frac{m_u}{q_k + b_k + A_k} \tag{19}$$

In this relation, $m_u$ represents the number of qubits transmitted, $q_k$ is the number of qubits used as the channel, $b_k$ is the number of classical bits transmitted, and $A_k$ represents the number of auxiliary qubits used. Thus, we have:

$$m_u = 2n, q_k = 4n+1, b_k = 0, A_k = n, \eta = \frac{2n}{6n+1} \qquad (20)$$

As obtained from the above relationship, we can see that as the number of n increases, the efficiency of this protocol will be closer to the value of one. There is a comparison in Table 2 between previous researches and the proposed scheme. The efficiency of some of these calculations has been computed for a 6-qubit system.

*Table 2-Compasrison between previous studies and proposed scheme*

| Scheme | Number of transferred qubits | Quantum channel | Classical channel | Efficiency |
|---|---|---|---|---|
| BQT[25] | 2n | 4n | 0 | 50% |
| BQT[21] | 2n | 4 qubits cluster state | 0 | 66% |
| BQT[19] | 2n | 2n qubits Bell state | Needed | 33.33% |
| QT[18] | n | 3n qubits | 0 | 33.33% |
| BQT[20] | 2n | (2n+2) qubits entangled state | 0 | 40% |
| Proposed scheme | 2n | (4n+1) qubits entangled state | 0 | 33.33% |

## 4. Conclusion

The scheme examined in this study, being a combination of quantum teleportation and remote state preparation, may initially seem relatively complex in terms of implementation, advancements in quantum hardware design will reduce this complexity over time. On the other hand, due to the relatively high efficiency of this protocol, especially with an increasing number of qubits, its application becomes more logical in systems requiring the transfer of a larger number of qubits. Additionally, as mentioned earlier, this protocol allows both the sender and receiver to simultaneously transmit their quantum information to one another, making it a higher priority for systems that require such functionality.